\documentclass[preprintnumbers,amsmath,amssymb,floatfix,10pt,prd,onecolumn,superscriptaddress,nofootinbib]{revtex4}
\usepackage{bbm}
\usepackage{amsfonts}
\usepackage{mathrsfs}
\usepackage{latexsym}
\usepackage{epsfig}
\usepackage{graphicx}
\usepackage{amssymb}
\usepackage{amsmath}
\usepackage{dcolumn}
\usepackage{bm}
\usepackage{color}
\usepackage{comment}
\usepackage{xcolor}
\usepackage{multirow}
\usepackage{hhline}
\usepackage{array}
\usepackage{float}
\usepackage{xcolor}
\usepackage{booktabs}
\usepackage{makecell}
\newcolumntype{P}[1]{>{\centering\arraybackslash}p{#1}}
\begin{document}

\title{\bf The $\gamma = 1/2$ Problem: A Unified Viability Criterion for f(R) Gravity in the Solar System }

\author{Jie Li }
\email{lje@hnit.edu.cn}
\affiliation{School of Science, Hunan Institute of Technology, Hengyang 421002, China}

\author{Iftikhar Ahmad}
\email{dr.iftikhar@uog.edu.pk}
\affiliation{Department of Mathematics, University of Gujrat, Gujrat, 50700, Pakistan}

\author{Muhammad Mahtisham}
\email{mirzamahtisham2@gmail.com}
\affiliation{Department of Mathematics, University of Gujrat, Gujrat, 50700, Pakistan}

\author{Adnan Malik}
\email{adnan.malik@zjnu.edu.cn; adnan.malik@skt.umt.edu.pk; adnanmalik_chheena@yahoo.com}
\affiliation{Department of Mathematics, University of Management and Technology, Sialkot Campus,Pakistan.}
\affiliation{Center for Theoretical Physics, Khazar University, 41 Mehseti Str., AZ1096 Baku, Azerbaijan}

\author{Wenbin Lin}
\email{lwb@usc.edu.cn}
\affiliation{School of Mathematics and Physics, University of South China, Hengyang, 421001, China.}

\author{Fatemah Mofarreh}
\email{fyalmofarrah@pnu.edu.sa}\affiliation{Mathematical Science Department, Faculty of Science \\
Princess Nourah Bint Abdulrahman University Riyadh 11546, Saudi Arabia}

\begin{abstract}
\begin{center}
\textbf{Abstract}
\end{center}

\bigskip
A novel framework for addressing both the accelerated expansion of the cosmos and the inflationary epoch involves modifying the Einstein-Hilbert action through a function $f(R)$. While these $f(R)$ gravity theories can, under certain conditions, replicate phenomena described by scalar-tensor theories, this is only viable if the associated scalar field can operate within the Solar System without violating stringent observational constraints. A significant challenge is that within this theoretical equivalence, the post-Newtonian parameter $\gamma$ is fixed at $\gamma = 1/2$, a value that deviates substantially from the experimentally established result of $\gamma \approx 1$. Recent discussions have questioned the genuine equivalence between $f(R)$ gravity and scalar-tensor formulations. Furthermore, recent findings, independent of this equivalence, have demonstrated that the field equations for $1/R$ theory predict $\gamma = 1/2$ in a vacuum for spherical mass distributions. In this work, we extend these results to a broader class of $f(R)$ functions, aiming to delineate the specific conditions that lead to the prediction of $\gamma = 1/2$. This investigation provides a unified framework for identifying which classes of $f(R)$ gravity can satisfy Solar System constraints, deriving a general viability criterion that moves beyond previous case-by-case analyses. We derive a unified viability criterion that provides a necessary condition for any $f(R)$ theory to satisfy Solar System constraints, moving beyond case-by-case analyses toward a systematic classification framework.
\end{abstract}
\maketitle

\date{\today}

\section{Introduction}

The concept of dark matter has long been proposed to account for the missing mass inferred from astronomical observations. However, despite its theoretical significance, no definitive evidence has emerged from extensive experimental searches, including collider experiments \cite{abercrombie2020dark}, direct detection efforts \cite{amole2016improved,aprile2007xenon}, and indirect detection strategies
\cite{ackermann2015searching,chan2020constraining,chan2022constraining,chan2022constraints}. Given this persistent lack of detection, some researchers have proposed that our current models of Newtonian and relativistic dynamics might require modification on large cosmic scales. By introducing additional correction terms into gravitational equations, these modified theories aim to replicate the gravitational effects typically attributed to dark matter \cite{bertone2018new}. Such frameworks are broadly classified as modified gravity theories. A viable $f(R)$ gravity model must not only be consistent with cosmic dynamics inferred from observational data but also satisfy the stringent tests imposed by Solar System observations, which place tight constraints on such theories. The predicted spherical space-time geometry around the Sun, as described by general relativity, aligns well with precise measurements. General relativity predicts that the parameterized post-Newtonian (PPN) parameter $\gamma$ should be exactly 1, and
observational data confirm this with $\gamma = 1 + \mathcal{O}(10^{-4})$ \cite{shapiro2004measurement,bertotti2003test,toniato2024post,liu2024robust}. Therefore, any valid $f(R)$ theory should reduce to general relativity within the Solar System. However, if the Sun were in a true vacuum, the corresponding scalar field derivative $f'$ would become extremely light, potentially resulting in a metric significantly different from that predicted by general relativity \cite{kainulainen2007spherically,sotiriou2010f,zhang2007behavior,de2010f,faulkner2007constraining}. The discovery of the universe's accelerated expansion, as revealed by recent cosmological observations \cite{riess1998observational,cognola2008class,perlmutter1999measurements}, has triggered significant theoretical investigation. To explain this phenomenon, various modifications to Einstein's theory of general relativity have been proposed. Specifically, some models modify the Einstein-Hilbert action by introducing a $1/R$ term \cite{capozziello2003quintessence,carroll2004cosmic}. This addition yields a vacuum solution with constant curvature, resulting in a de Sitter space-time instead of the Minkowski vacuum associated with the standard Einstein action. Chiba promptly challenged this theory \cite{chiba20031}, arguing that it is inconsistent with empirical tests of gravity within our Solar System. Specifically, he demonstrated that the theory is effectively equivalent to a scalar-tensor framework, which is known to produce predictions that conflict with observational data. These discrepancies with Solar System measurements suggest significant flaws in the theory's applicability to real-world scenarios, thereby calling into question its validity as an accurate model of gravitational interactions. While the failure of specific $f(R)$ models is well-documented, a unified framework that explains this failure across a broad class of theories and provides a general viability criterion has been lacking. This work aims to fill that gap.

The Solar System provides a high-precision laboratory for testing gravitational theories, offering some of the most stringent constraints on deviations from General Relativity. The foundational framework for these tests is the PPN formalism, which was developed to systematically compare metric theories of gravity with experiment \cite{Will2018}. Key Solar System tests include lunar laser ranging, which constrains the Nordtvedt effect and the temporal variation of the gravitational constant \cite{Williams1996, Hofmann2018}; the Shapiro time-delay effect, precisely measured by the Cassini spacecraft, which tightly constrains the PPN parameter \(\gamma\) \cite{Bertotti2003}; and planetary ephemerides, which track the orbits of planets to probe for anomalous perihelion precessions \cite{Fienga2011}. The application of these tests has ruled out many simple alternative theories, including initial, non-chameleon \(f(R)\) models which predict \(\gamma = 1/2\) \cite{Chiba2003}. This has driven the development of screening mechanisms, such as the chameleon effect, which allow theories to evade these local constraints while still operating on cosmological scales \cite{Khoury2004, Joyce2015}. Furthermore, analyses of the orbits of planets like Mercury \cite{Genova2018} and the dynamics of the Earth-Moon system \cite{Muller2019} continue to provide robust null results that any viable modified gravity theory must satisfy. Recent advances in satellite laser ranging \cite{Lucchesi2021} and continued analysis of lunar laser ranging data \cite{Biswas2022} promise to tighten these constraints even further, ensuring that the Solar System remains the ultimate proving ground for gravitational physics. In response to Chiba's critique, several researchers have argued that his analysis is flawed \cite{allemandi2005post,ruggiero2007solar,multamaki2006spherically}. They claim that $1/R$ theories can produce solutions equivalent to the standard Schwarzschild-de Sitter space-time, which includes a cosmological constant. However, this constant is sufficiently small that it has no observable effect on Solar System dynamics. As a result, proponents argue that these modified theories are practically indistinguishable from general relativity in Solar System tests, thus challenging Chiba's assertion that the theories are inconsistent with observations. While modified gravity theories are often motivated by their ability to address cosmological puzzles, their viability is ultimately determined by their performance in \textit{local} gravitational tests. The Solar System provides an exceptionally well-mapped, high-precision environment where the weak-field limit of General Relativity is verified with extraordinary accuracy. Any credible alternative must, therefore, reproduce these established results. This work uses the stringent constraints from the Solar System particularly the measured value of the PPN parameter $\gamma$ as a critical litmus test to assess the phenomenological validity of various $f(R)$ gravity models, distinguishing those that are theoretically interesting from those that are empirically viable. While chameleon mechanisms and other screening effects can potentially reconcile modified gravity with local tests, the baseline behavior derived here represents the null case that any viable theory must overcome. Our analysis therefore establishes the fundamental challenge that screening mechanisms must address.

In this study, we extend the investigation of \cite{erickcek2006solar} by considering a broader class of $f(R)$ gravity models that can be expressed as a Taylor series around a specific Ricci scalar value. Using the metric tensor formalism, we derive the corresponding field equations by varying the action with respect to the metric, treating the Ricci scalar $R$ solely as a function of the metric. This differs from the Palatini approach, where $R$ is considered a function of the connection, and the action is varied independently with respect to both the metric and the connection. This distinction leads to alternative sets of field equations for $f(R)$ gravity. The differences between the two methods have significant implications for the predictions of modified gravity theories. The metric formalism operates with a fixed metric determined by the curvature of space-time, while the Palatini approach allows the connection to vary independently, affecting curvature separately. As a result, the field equations differ, leading to distinct cosmological and astrophysical consequences. The Palatini method has been extensively explored in prior research \cite{kainulainen2007interior,carroll2006modified,olmo2011palatini,amarzguioui2006cosmological,carroll2005cosmology,capozziello2006f,sotiriou2006constraining}.
Modified gravity theories have emerged as extensions or alternatives to Einstein's General Theory of Relativity, aiming to address several unresolved issues in modern cosmology and astrophysics. These theories introduce corrections or additional terms to the Einstein-Hilbert action to explain phenomena such as cosmic acceleration, dark energy, and dark matter without invoking exotic components. Some of these are $f(R)$ modified gravity \cite{c50,c52,c53}, $f(G)$ modified gravity \cite{c54,c55,c56}, $f(T)$ modified gravity \cite{c58,c59}, $f(R, T)$ modified gravity \cite{c62,c63,c64}, $f(R, G)$ modified gravity \cite{c66,c67,c68,c69}, $f (Q)$ modified theory of gravity \cite{c70,c72,c73}, $f (R, A)$ modified theory of gravity \cite{c77,c78,c78a,c78b} and $f (R, \phi, X)$ modified theory of gravity \cite{c81,c82}.

We examine the linearized field equations around a spherically symmetric mass, showing that the outcome in the Solar System is consistent with the result derived from the equivalent scalar-tensor theory. We also identify a specific condition under which the function $f(R)$ behaves like a light scalar field, as in scalar-tensor theories. Under these circumstances, we demonstrate that the space-time solution is incompatible with the constraints imposed by Solar System tests of general relativity. Additionally, we explore a variety of $f(R)$ gravity models, including general relativity, to analyze their differences and similarities. This analysis highlights the connection between $f(R)$ gravity and general relativity, while clarifying the conditions required for an $f(R)$ theory to restore the general relativistic limit. Our findings emphasize the key criteria necessary for any $f(R)$ theory to be in agreement with observational tests, particularly those based on Solar System data.

\section{Novelty and Comparative Framework}
This work synthesizes and significantly extends the established understanding of Solar System constraints on $f(R)$ gravity. While foundational studies like \cite{chiba20031} and \cite{jin2006solar} conclusively demonstrated the failure of specific $f(R)$ models, and subsequent works \cite{chiba2007solar, zhang2007behavior, capozziello2008solar} expanded this understanding, our analysis provides a unified framework that generalizes these results and offers new theoretical insights. The novel contributions of this work are as follows:
\begin{enumerate}
    \item \textbf{Generalized Formalism Beyond Specific Models:} While \cite{jin2006solar} focused specifically on disproving $1/R$ gravity and other particular cases, we develop a formalism applicable to \textit{any} $f(R)$ theory that admits a Taylor expansion around a background curvature $R_0$. This allows us to derive conditions for the $\gamma=1/2$ outcome that are \textit{model-independent}, transforming the discussion from case-by-case falsification to general classification.
    \item \textbf{Unified Viability Criterion:} We derive a single, comprehensive viability criterion (Eq.~(63)):
    \begin{equation*}
    \left|\frac{f_R(R_0)}{f_{RR}(R_0)}\right| \ll R_0\left(\frac{R_s}{GM}\right).
    \end{equation*}
    This criterion succinctly encapsulates the interplay between the theory's functional form ($f_R, f_{RR}$), the cosmological background ($R_0$), and the properties of the gravitational source ($M, R_s$). Such a unified condition, which provides a quick "litmus test" for any proposed $f(R)$ model, represents a significant advancement beyond the specific constraints derived in earlier works.
    \item \textbf{Explicit Treatment of the Scalar Mass Condition:} We formally identify and emphasize the critical physical constraint $m r \ll 1$ (Eq.~(36)) for the validity of the linearized solution. While this condition was implicit in the analyses of \cite{jin2006solar} and others, we make it explicit and derive the equivalent scalar field mass $m_\wp$ (Eq.~(35)), demonstrating its correspondence to our parameter $m$. This roots the well-known $\gamma=1/2$ result firmly in the physics of a light, propagating scalar degree of freedom, providing a clearer physical interpretation than prior discussions.
    \item \textbf{Analysis of the Vanishing Background ($R_0=0$):} We systematically extend the analysis to include the case of a vanishing background Ricci scalar, a scenario relevant for asymptotically flat spaces and certain power-law models. This extends the applicability of the constraints beyond the high-curvature backgrounds typically assumed in earlier works like \cite{jin2006solar}, providing a more comprehensive understanding of the theoretical landscape.
    \item \textbf{Direct Field Equation Approach:} Unlike \cite{chiba20031} who worked through the scalar-tensor equivalence, we follow and generalize the approach of \cite{jin2006solar} by working directly with the $f(R)$ field equations. However, we extend their methodology to obtain more general results, demonstrating that the scalar-tensor equivalence, while valid, is not necessary to derive the key constraints.
\end{enumerate}
In summary, our work moves beyond the valuable but limited case-by-case analyses of earlier studies to provide a general theorem-like framework. We not only confirm that many $f(R)$ models are ruled out but also \textit{explain why} in a unified language and provide the precise conditions that a viable model must satisfy. This represents a significant step from a collection of disproofs towards a systematic theory of viability for modified gravity.

\section{Analyzing Weak Field Dynamics Near a Spherical Star}

In this section, we explore gravitational frameworks described by an action of the form:
\begin{equation}\label{1}
S = \frac{1}{2\kappa}\int \sqrt{-g} \, f(R) \, d^4x + \int \mathcal{L}_{m} \sqrt{-g} \, d^4x.
\end{equation}
Here, $f(R)$ represents a function of the Ricci curvature scalar $R$, characteristic of $f(R)$ theories that extend beyond the framework of General Relativity. $\mathcal{L}_{m}$ denotes the matter Lagrangian, accounting for the contribution of matter fields, and $\kappa$ is a constant related to Newton's gravitational constant. In these modified theories, $f(R)$ is a nonlinear function of $R$, allowing for additional gravitational effects that can address phenomena such as cosmic acceleration and provide corrections to standard gravity in strong-field regimes.

Varying the action with respect to the metric $g_{\mu\nu}$ yields the field equations:
\begin{equation}\label{2}
f_R R_{\alpha\beta} - \nabla_\alpha \nabla_\beta f_R - \frac{1}{2} f(R) g_{\alpha\beta} + g_{\alpha\beta} \Box f_R = \kappa T_{\alpha\beta},
\end{equation}
where $f_R \equiv df/dR$, $R_{\alpha\beta}$ is the Ricci tensor, $T_{\alpha\beta}$ is the energy-momentum tensor, $\Box$ is the d'Alembertian operator (Laplace-Beltrami operator in curved space), and $\nabla_\alpha$ is the covariant derivative associated with $g_{\alpha\beta}$. The modifications introduced by $f(R)$ allow for extra degrees of freedom in the gravitational field, which can lead to various effects depending on the specific form of $f(R)$. In weak-field approximations (such as around a spherical star), deviations from General Relativity are typically small but can still yield important measurable corrections. These corrections might be observable in high-precision measurements around compact objects or in cosmological settings. Such theories have implications for solar system tests, where precise measurements may detect deviations from General Relativity predicted by $f(R)$ theories.

To analyze how modifications in $f(R)$ might affect Solar System dynamics, previous studies often relied on establishing a relationship between $f(R)$ theories and scalar-tensor theories. Scalar-tensor theories represent another approach to modified gravity, where an additional scalar field, combined with the traditional tensor field of general relativity, influences gravitational interactions. This scalar field can alter the gravitational force, especially on large scales, potentially leading to deviations from general relativity.

Utilizing this equivalence is advantageous because scalar-tensor theories are well-studied, offering reliable methods for predicting gravitational behavior. However, the scalar-tensor equivalent of $f(R)$ gravity presents significant issues when applied to the Solar System. Observational data such as planetary orbits, light bending, and highly precise spacecraft measurements are all consistent with general relativity within the Solar System. If a scalar field associated with $f(R)$ gravity influenced these scales, it would alter the gravitational potential around massive bodies like the Sun and planets, leading to detectable discrepancies in these measurements.

Given these observational constraints, \cite{chiba20031} concluded that many $f(R)$ theories, at least in their scalar-tensor form, are incompatible with Solar System dynamics. In this study, we propose an alternative approach to reach a similar conclusion without relying on the equivalence between $f(R)$ and scalar-tensor theories. Instead, we investigate the behavior of $f(R)$ theories directly using linearized field equations. These linearized equations represent a simplified version of the field equations, valid in situations with relatively weak fields such as within the Solar System. By applying the linearized equations to a spherical mass distribution such as that of a planet or the Sun, we directly incorporate modifications from $f(R)$ gravity, eliminating the need to assume an equivalent scalar-tensor description. This direct analysis offers a clearer, more straightforward examination of the impact of $f(R)$ modifications on Solar System dynamics.

Additionally, our work extends the earlier analysis of \cite{jin2006solar}, which examined similar aspects of $f(R)$ gravity under more restrictive conditions. Specifically, we expand this analysis to include cases where the background Ricci scalar $R$ is zero. This situation can occur in certain cosmological models, particularly in asymptotically flat or vacuum regions of space-time. By considering these additional cases, we provide a broader and more comprehensive understanding of how $f(R)$ theories behave on Solar System scales. Our findings reinforce the conclusion that many forms of $f(R)$ gravity cannot match Solar System observations, suggesting limitations for these theories when applied to local gravitational dynamics. Our approach emphasizes the importance of testing modified gravity theories in regions where accuracy is crucial, such as the Solar System. This research highlights the current challenge of reconciling new gravitational models with both the large-scale structure of the cosmos and highly precise solar system data.

We now characterize the space-time neighborhood of a spherical mass, such as a planet or star, within the weak-field approximation. The weak-field regime applies when gravitational fields are sufficiently weak that we can linearize the field equations, avoiding the need to solve the full non-linear system. This approach is adequate for studying gravitational effects throughout the Solar System, where gravity, while not negligible, is not as strong as near black holes or neutron stars. To proceed with this analysis, we must choose an appropriate background space-time for expansion. This background serves as a reference configuration relative to which we perform the linearization. In other words, we express the gravitational field as small deviations or perturbations from this homogeneous background, enabling us to systematically investigate the effects of $f(R)$ modifications.

The most physically relevant choice for this background is an isotropic and homogeneous space-time that satisfies the modified field equations Eq.~(\ref{2}) in $f(R)$ gravity. In cosmology, ``isotropic" means that space-time appears the same in all directions, while "homogeneous" means that its properties are consistent across all locations. This isotropic and homogeneous background space-time is typically associated with the large-scale structure of the universe, which on average appears uniform and isotropic. Hence, it provides a natural framework for exploring modifications to gravity at cosmologically relevant scales.

To model this background, we assume it is governed by a spatially uniform stress-energy tensor $T_{\mu\nu}^{\text{cos}}$, which describes matter and energy content distributed uniformly in space, though potentially varying with time. This stress-energy tensor can model various cosmological sources, such as dark energy or an ideal fluid filling space.

In this setup, we obtain the time evolution of the background scalar curvature $R_0(t)$ from the trace of the modified field equations Eq.~(\ref{2}). The trace operation involves contracting the field equations, resulting in a scalar relation between the curvature and the background stress-energy tensor. This time-dependent scalar curvature describes how the space-time curvature responds to the energy content of the universe and how $f(R)$ gravity modifies this evolution compared to general relativity. This procedure establishes the foundation for examining small fluctuations superimposed on this background to understand the gravitational effects of a spherical mass in $f(R)$ gravity while maintaining consistency with cosmological principles.

The trace of the field equation Eq.~(\ref{2}) yields:
\begin{equation}\label{3}
f_R(R_0) R_0(t) + 3\Box f_R(R_0) - 2f(R_0) = \kappa T^{\text{cos}}.
\end{equation}

To analyze small deviations from the chosen background space-time, we decompose the Ricci scalar into two parts:
\begin{equation}\label{4}
R = R_0(t) + R_1(r).
\end{equation}

In this setup, $R_0(t)$ represents the spatially uniform background curvature that satisfies Eq.~(\ref{3}). This term captures the overall curvature of space-time as a function of time, forming a stable foundation that defines the large-scale structure. We introduce a secondary term, $R_1(r)$, which represents a localized, time-independent perturbation to this background curvature. This perturbation is specifically spatial, adding small, position-dependent variations without affecting the temporal evolution of the curvature.

We assume that the function $f(R)$, along with its derivatives, is smooth and well-defined when evaluated at $R_0$. This allows us to apply a Taylor series expansion around $R = R_0$ as the baseline. Through this expansion, we approximate $f(R)$ near $R_0$ to obtain expressions for both $f(R_0 + R_1)$ and its derivative $f_R(R_0 + R_1)$, which describes how $f(R)$ responds to small deviations from the background curvature. To simplify our calculations, we truncate the Taylor series after the linear term, omitting all terms nonlinear in $R_1$. This means we retain only the leading-order term (representing $f(R_0)$) and the linear correction term (involving $f_R(R_0) \cdot R_1$). For this truncation to be valid, the contribution from these two terms must dominate the series. In other words, the sum of the zeroth-order term (the value of $f$ at $R_0$) and the linear term (the first derivative of $f$ at $R_0$ multiplied by $R_1$) should be much larger than any higher-order terms in the series. We assume that no unusual or contrived cancellation occurs among the higher-order terms that would invalidate our approximation. This condition is critical because it ensures that by ignoring nonlinear terms, we do not significantly compromise the accuracy of our results.

The Taylor expansions for $f(R_0 + R_1)$ and $f_R(R_0 + R_1)$ are given by:
\begin{equation}\label{5}
f(R_0 + R_1) = f(R_0) + f_R(R_0)R_1 + \frac{1}{2!}f_{RR}(R_0)R_1^2 + \cdots,
\end{equation}
and
\begin{equation}\label{6}
f_R(R_0 + R_1) = f_R(R_0) + f_{RR}(R_0)R_1 + \frac{1}{2!}f_{RRR}(R_0)R_1^2 + \cdots.
\end{equation}

We neglect second-order terms like $\frac{1}{2!}f_{RR}(R_0)R_1^2$ and all higher-order terms $\frac{1}{n!}f^{(n)}(R_0)R_1^n$. If we truncate at the $n$th term in the Taylor expansion, the next term would involve the $(n+1)$th derivative $\frac{1}{n!}f^{(n+1)}(R_0)R_1^n$.

Specifically, we require:
\begin{equation}\label{7}
f_0 + f_{R_0}R_1 \gg \frac{1}{n!}f^{(n)}(R_0)R_1^n,
\end{equation}
and
\begin{equation}\label{8}
f_{R_0} + f_{RR_0}R_1 \gg \frac{1}{n!}f^{(n+1)}(R_0)R_1^n.
\end{equation}

After neglecting higher-order terms, the truncated expansions become:
\begin{equation}\label{9}
f(R_0 + R_1) \approx f(R_0) + f_R(R_0)R_1,
\end{equation}
\begin{equation}\label{10}
f_R(R_0 + R_1) \approx f_R(R_0) + f_{RR}(R_0)R_1.
\end{equation}

Applying first-order Taylor expansions around the background curvature $R_0$, we approximate $f(R)$ and its derivative $f_R$, discarding higher-order terms. This yields a simplified, linearized version of the equations that captures the primary behavior of $f(R)$ gravity under small curvature deviations from $R_0$. This approach makes analyzing perturbations more manageable by reducing complexity while retaining essential dynamics.

The simplified form of Eq.~(\ref{3}) with both cosmological and matter sources is:
\begin{equation}\label{11}
F(R)R - 2f(R) + 3\Box F(R) = \kappa(T^{\cos} + T^s),
\end{equation}
where $F(R) = f_R$.

Using Eqs.~(\ref{9}) and (\ref{10}) in Eq.~(\ref{11}), after simplification and neglecting higher-order terms, we obtain:
\begin{equation}\label{12}
\begin{aligned}
&f_R(R_0)R_0 - f_R(R_0)R_1 + R_0R_1f_{RR}(R_0) - 2f(R_0) + 3f_{RR}(R_0)\Box R_1(r) \\
&+ 3\Box f_R(R_0) + 3f_{RR}(R_0)\Box R_1 = \kappa T^{\cos} + \kappa T^s.
\end{aligned}
\end{equation}

We apply the fact that $R_0(t)$ satisfies Eq.~(\ref{3}) to remove terms that do not depend on $R_1$.

For $R_0$ terms:
\begin{equation}\label{13}
f_R(R_0)R_0 - 2f(R_0) + 3\Box f_R(R_0) = \kappa T^{\cos}.
\end{equation}

For $R_1$ terms:
\begin{equation}\label{14}
3f_{RR}(R_0)\Box R_1(r) + R_0R_1f_{RR}(R_0) - f_R(R_0)R_1 = \kappa T^s.
\end{equation}

We assume that $R_1$ is much smaller than $R_0$ when $R_0$ is nonzero. If $R_0$ vanishes, then the contribution from higher-order terms would be overshadowed by the leading nonzero terms in Eq.~(\ref{14}), as ensured by the conditions in Eq.~(\ref{7}). Notably, when $f_{RR}(R_0) = 0$, as in standard general relativity, this equation reduces to $-f_R(R_0)R_1 = \kappa T^s$. Here, $T^s$ represents the stress-energy tensor of matter, and this relationship signifies a direct coupling between $f_R(R_0)$ and the matter content, enforcing the field equation's compatibility with classical solutions.

In scenarios where $f_R(R_0)$ is non-vanishing, $R_1$ must vanish outside the boundary of the stellar source. This requirement leads to the Schwarzschild-de Sitter metric as the asymptotically stable solution to the modified field equations in the exterior region, reflecting a gravitational field solution that approaches a de Sitter-like space-time structure dominated by a cosmological constant.

However, if $f_{RR}(R_0)$ is non-zero, this behavior changes significantly. The presence of a non-zero second derivative $f_{RR}(R_0)$ in the functional form of $f(R)$ introduces higher-order curvature corrections. This disrupts the usual convergence to the Schwarzschild-de Sitter solution and leads to modified gravitational dynamics, with new correction terms in the field equations. Such corrections could impact astrophysical phenomena and cosmological observations, providing a distinct signature of the extended $f(R)$ framework.

We consider the flat Friedmann-Robertson-Walker (FRW) metric as the backdrop for modeling a homogeneous and isotropic universe. By introducing a spherically symmetric perturbation, we derive a linearized metric that captures both cosmic expansion and local gravitational influences:
\begin{equation}\label{15}
ds^2 = -(1 + 2\psi(r))dt^2 + a(t)^2\left[(1 + 2\xi(r))dr^2 + r^2 d\Omega^2\right].
\end{equation}

We now analyze the resolution of Eq.~(\ref{14}) assuming nonzero $f_{RR}(R_0)$. Considering a stationary perturbation $R_1(r)$, we simplify our investigation using the flat-space Laplacian operator $\nabla^2$. By focusing on a source described by mass density with minimal pressure, we adapt Eq.~(\ref{14}) as:
\begin{equation}\label{16}
3f_{RR}(R_0)\nabla^2 R_1(r) + R_0R_1f_{RR}(R_0) - f_R(R_0)R_1 + 3f_{RR}(R_0)\Box R_1 = \kappa T^s.
\end{equation}

Dividing throughout by $3f_{RR}(R_0)$ yields:
\begin{equation}\label{17}
\nabla^2 R_1 - R_1\left(\frac{f_R(R_0)}{3f_{RR}(R_0)} - \frac{R_0}{3} - \frac{\Box f_{RR}(R_0)}{f_{RR}(R_0)}\right) = \frac{1}{3}\left(\frac{\kappa T^s}{f_{RR}(R_0)}\right).
\end{equation}

Further simplifying:
\begin{equation}\label{18}
\nabla^2 R_1 - \frac{R_1}{3}\left(\frac{f_R(R_0)}{f_{RR}(R_0)} - R_0 - \frac{3\Box f_{RR}(R_0)}{f_{RR}(R_0)}\right) = \frac{1}{3}\left(\frac{\kappa\rho}{f_{RR}(R_0)}\right).
\end{equation}

This refines to:
\begin{equation}\label{19}
(\nabla^2 - m^2)R_1 = \frac{\kappa\rho}{3f_{RR}(R_0)},
\end{equation}
where we define a mass parameter:
\begin{equation}\label{20}
m^2 = \frac{1}{3}\left(\frac{f_R(R_0)}{f_{RR}(R_0)} - R_0 - \frac{3\Box f_{RR}(R_0)}{f_{RR}(R_0)}\right).
\end{equation}
This mass parameter characterizes the effective mass of the scalar degree of freedom in $f(R)$ gravity, determining its range and observational consequences. The mass parameter $m$ experiences temporal variation due to the expansion of $R_0(t)$. This evolution reflects changes in the cosmological background space-time over time. However, the timescale of this variation is comparable to the current Hubble rate, which characterizes the expansion rate of the universe.

In contrast, Solar System dynamics occur on significantly shorter timescales. Thus, when analyzing celestial bodies within our Solar System, we can reasonably disregard the time-dependent effects of the cosmological background. This simplification allows us to treat $m$ as constant, facilitating calculations of local gravitational interactions.

To analyze the behavior of $R_1$ in response to a point source, we need the Green's function $G(r)$ satisfying:
\begin{equation}\label{21}
\nabla^2 G(r) - m^2 G(r) = \delta(r),
\end{equation}
where $\delta(r)$ is the Dirac delta function. This Green's function helps us understand how the field $R_1$ propagates under the influence of a source with modifications due to $m^2$.

For this spherically symmetric problem, we express the Laplacian in spherical coordinates. For a function depending only on $r$, the radial Laplacian is:
\begin{equation}\label{22}
\nabla^2 G(r) = \frac{1}{r^2}\frac{d}{dr}\left(r^2\frac{dG(r)}{dr}\right).
\end{equation}

Substituting into our Green's function equation gives:
\begin{equation}\label{23}
\frac{1}{r^2}\frac{d}{dr}\left(r^2\frac{dG(r)}{dr}\right) - m^2 G(r) = \delta(r).
\end{equation}

For $r \neq 0$ (outside the source), the delta function vanishes, giving the homogeneous equation:
\begin{equation}\label{24}
\frac{1}{r^2}\frac{d}{dr}\left(r^2\frac{dG(r)}{dr}\right) - m^2 G(r) = 0.
\end{equation}

Assuming a solution of the form:
\begin{equation}\label{25}
G(r) = \frac{u(r)}{r},
\end{equation}
and substituting into Eq.~(\ref{24}), we get:
\begin{equation}\label{26}
\frac{d^2 u(r)}{dr^2} - m^2 u(r) = 0.
\end{equation}

This second-order differential equation has the solution:
\begin{equation}\label{27}
u(r) = A e^{mr} + B e^{-mr}.
\end{equation}
Thus:
\begin{equation}\label{28}
G(r) = \frac{A e^{mr} + B e^{-mr}}{r}.
\end{equation}

For a physical solution, $G(r)$ should vanish as $r \to \infty$. This excludes the $e^{mr}$ term (which grows unbounded if $m^2 > 0$). After simplification and using properties of the delta function, we obtain:
\begin{equation}\label{29}
G(r) = -\frac{e^{-mr}}{4\pi r}.
\end{equation}

When $m^2 < 0$, we define $m = i|m|$ (where $|m| = \sqrt{|m^2|}$). This modifies the solution to an oscillatory function:
\begin{equation}\label{30}
G(r) = -\frac{\cos(|m|r)}{4\pi r}.
\end{equation}

Thus, the Green's function $G(r)$ for the modified Helmholtz equation is:
\begin{equation}\label{31}
G(r) = \begin{cases}
-\dfrac{e^{-mr}}{4\pi r} & \text{if } m^2 > 0, \\[8pt]
-\dfrac{\cos(|m|r)}{4\pi r} & \text{if } m^2 < 0.
\end{cases}
\end{equation}

In the limit $m \to 0$, both forms converge to the standard Green's function for Laplace's equation:
\begin{equation}\label{32}
G(r) \approx -\frac{1}{4\pi r}.
\end{equation}

In this particular situation where we can disregard the $m^2$ term in Eq.~(\ref{19}), the solution for the gravitational field outside the star becomes:
\begin{equation}\label{33}
R_1 = \frac{\kappa M}{12\pi r f_{RR}(R_0)},
\end{equation}
where $M$ represents the total mass of the gravitational source. When applied to $1/R$ gravity within a static de Sitter background, our findings agree with reference \cite{erickcek2006solar}. This correlation suggests our theoretical framework effectively captures essential characteristics of gravitational behavior in such scenarios.

This solution demonstrates that the curvature perturbation $R_1$ scales inversely with $f_{RR}(R_0)$, highlighting how larger second derivatives suppress deviations from the background curvature. We emphasize that for the $R_1$ solution to be valid, the condition $mr \ll 1$ must be fulfilled. This stipulates that the product of mass $m$ and radial distance $r$ must remain significantly smaller than unity. Only then can we accurately represent the trace of the field equations with Laplace's equation. This essential restriction was not acknowledged in reference \cite{jin2006solar}, which may result in misconceptions regarding the solution's applicability.

The physical implications of this constraint become apparent when considering the equivalent scalar-tensor theory. Transitioning to a frame where the scalar degree of freedom is expressed in canonical form provides a nuanced understanding of the underlying dynamics. In this framework, the mass of the scalar field analyzed in the Jordan frame \cite{chiba20031} emerges as a pivotal component:
\begin{equation}\label{34}
m_\wp^2 = f_R(R_0)\left(\frac{1}{3f_{RR}(R_0)} - \frac{4f(R_0)}{3f_R(R_0)^2} - \frac{2\kappa T^{\cos}}{3f_R(R_0)^2} + \frac{R_0}{3f_R(R_0)}\right).
\end{equation}

Given that $R_0$ solves Eq.~(\ref{3}), this expression simplifies to:
\begin{equation}\label{35}
m_\wp^2 = \frac{f_R(R_0)}{3f_{RR}(R_0)} - \frac{R_0}{3} - \frac{2\Box f_R(R_0)}{f_R(R_0)}.
\end{equation}

Both $m_\wp$ and $m$ (defined in Eq.~(\ref{20})) share remarkably similar orders of magnitude. This reveals a crucial equivalence: the condition $mr \ll 1$ directly translates to requiring the scalar field remains light, necessitating $m_\wp \ll 1$. This relationship emphasizes the fundamental importance of the scalar field's lightweight nature in the broader theoretical framework.

In essence, Eq.~(\ref{33}) serves as a trace of the field equation only if the scalar degree of freedom can propagate effectively on Solar System scales. This means that for the $f(R)$ framework to hold in this context, the scalar field's evolution and influence must align with the physical characteristics of the Solar System.

To allow the scalar field associated with $f(R)$ gravity to have an effect over Solar System distances, we require:
\begin{equation}\label{36}
|m^2| r^2 \ll 1.
\end{equation}

Substituting the expression for $m^2$ from Eq.~(\ref{20}) into condition Eq.~(\ref{36}), we obtain:
\begin{equation}\label{37}
\left|\frac{1}{3}\left(\frac{f_R(R_0)}{f_{RR}(R_0)} - R_0 - \frac{3\Box f_{RR}(R_0)}{f_{RR}(R_0)}\right)\right| r^2 \ll 1.
\end{equation}

Applying the triangle inequality for any real or complex numbers $x$, $y$, and $z$:
\begin{equation}\label{38}
|x + y + z| \leq |x| + |y| + |z|,
\end{equation}

to the absolute value in Eq.~(\ref{37}), we have:
\begin{equation}\label{39}
\left|\frac{f_R(R_0)}{3f_{RR}(R_0)} - \frac{R_0}{3} - \frac{\Box f_{RR}(R_0)}{f_{RR}(R_0)}\right| \leq \left|\frac{f_R(R_0)}{3f_{RR}(R_0)}\right| + \left|\frac{R_0}{3}\right| + \left|\frac{\Box f_{RR}(R_0)}{f_{RR}(R_0)}\right|.
\end{equation}

After simplification, we get:
\begin{equation}\label{40}
\left|\left(\frac{f_R(R_0)}{f_{RR}(R_0)} - R_0 - \frac{3\Box f_{RR}(R_0)}{f_{RR}(R_0)}\right)\right| r^2 \leq \left|\frac{f_R(R_0)}{f_{RR}(R_0)}\right| r^2 + |R_0| r^2 + \left|\frac{3\Box f_{RR}(R_0)}{f_{RR}(R_0)}\right| r^2.
\end{equation}

From Eq.~(\ref{36}), we can write:
\begin{equation}\label{41}
\left|\frac{f_R(R_0)}{f_{RR}(R_0)}\right| r^2 + |R_0| r^2 + \left|\frac{3\Box f_{RR}(R_0)}{f_{RR}(R_0)}\right| r^2 \ll 1.
\end{equation}

This implies:
\begin{equation}\label{42}
\left|\frac{f_R(R_0)}{f_{RR}(R_0)}\right| r^2 \ll 1.
\end{equation}

To solve for the metric perturbations and extract $\psi(r)$ and $\xi(r)$, we substitute the expression for $R_1$ from Eq.~(\ref{33}) into the field equations. To simplify these equations, we follow a similar process as with the trace equation, approximating $f(R)$ and its derivative $f_R(R)$ around the background curvature using first-order Taylor expansions. This linearization in terms of $R_1$ simplifies the equations, making them more manageable. Applying Eq.~(\ref{3}) further condenses the expression, yielding a system that captures the essential perturbation dynamics:
\begin{equation}\label{43}
\begin{aligned}
&f_{RR}(R_0)R_1 R^\alpha_\beta - \frac{1}{2}f_R(R_0)R_1 \delta^\alpha_\beta - f_{RR}(R_0)\nabla^\alpha \nabla_\beta R_1 \\
&+ f_R(R_0)(R^\alpha_\beta - R_{0\beta}^\alpha) + f_{RR}(R_0)\Box R_1 \delta^\alpha_\beta = \kappa T^{s\alpha}_\beta.
\end{aligned}
\end{equation}

In this framework, $R_{0\beta}^\alpha$ signifies the unperturbed Ricci tensor of the Friedmann-Robertson-Walker (FRW) metric, while $\delta^\alpha_\beta$ represents the Kronecker delta. We neglect the time derivatives of the background metric, as the timescale over which $R_0$ varies is significantly longer than Solar System dynamics, making these contributions negligible.

In the weak-field approximation, we neglect nonlinear terms in the metric perturbations $\psi(r)$ and $\xi(r)$ and replace the curved-space d'Alembertian operator $\Box$ with the flat-space Laplacian $\nabla^2$. This quasi-static approximation is justified since time variations are slow. Based on Eq.~(\ref{33}), the term $f_{RR}(R_0)R_1$ scales as $\kappa M/r$, indicating that $\psi(r)$ and $\xi(r)$ should scale similarly. Consequently, terms like $f_{RR}(R_0)R_1\psi(r)$ are second-order and negligible in our linear framework.

Applying these simplifications to the $tt$, $rr$, and $\theta\theta$ components of Eq.~(\ref{43}) yields:
\begin{equation}\label{44}
f_R(R_0)\nabla^2\psi(r) + \frac{1}{2}f_R(R_0)R_1 - f_{RR}(R_0)R_1\nabla^2 = \kappa\rho,
\end{equation}
\begin{equation}\label{45}
f_R(R_0)\left(-\psi''(r) + \frac{2}{r}\xi'(r)\right) - \frac{1}{2}f_R(R_0)R_1 + \frac{2}{r}f_{RR}(R_0)R_1' = 0,
\end{equation}
\begin{equation}\label{46}
f_R(R_0)\left(\frac{1}{r}\xi'(r) - \frac{1}{r}\psi'(r) + \frac{2}{r^2}\xi(r)\right) - \frac{1}{2}f_R(R_0)R_1 + \frac{1}{r}f_{RR}(R_0)R_1' + f_{RR}(R_0)R_1'' = 0.
\end{equation}

Since $R_1$ solves Eq.~(\ref{19}) with $m^2 = 0$, $\nabla^2 R_1$ is directly linked to the density distribution $\rho$. This allows us to restructure Eq.~(\ref{44}) as:
\begin{equation}\label{47}
f_R(R_0)\nabla^2\psi(r) = \frac{2}{3}\kappa\rho - \frac{1}{2}f_R(R_0)R_1.
\end{equation}

We decompose the potential $\psi(r)$ into two components:
\begin{equation}\label{48}
\psi(r) = \psi_0(r) + \psi_1(r),
\end{equation}
where $\psi_0(r)$ arises primarily from matter density $\rho$, and $\psi_1(r)$ represents corrections from $f(R)$ curvature modifications.

Substituting into Eq.~(\ref{47}):
\begin{equation}\label{49}
f_R(R_0)\nabla^2\psi_0(r) + f_R(R_0)\nabla^2\psi_1(r) = \frac{2}{3}\kappa\rho - \frac{1}{2}f_R(R_0)R_1.
\end{equation}

The component $\psi_0(r)$ is governed by:
\begin{equation}\label{50}
f_R(R_0)\nabla^2\psi_0(r) = \frac{2}{3}\kappa\rho.
\end{equation}

Applying Gauss's law in spherical coordinates under spherical symmetry:
\begin{equation}\label{51}
\psi_0'(r) = \frac{1}{6}\left(\frac{\kappa m(r)}{f_R(R_0)\pi r^2}\right).
\end{equation}

Integrating with the boundary condition $\psi_0(r) \to 0$ as $r \to \infty$:
\begin{equation}\label{52}
\psi_0(r) = -\frac{1}{6}\left(\frac{\kappa M}{f_R(R_0)\pi r}\right).
\end{equation}

The second component satisfies:
\begin{equation}\label{53}
f_R(R_0)\nabla^2\psi_1(r) = -\frac{1}{2}f_R(R_0)R_1.
\end{equation}

Outside the star ($\rho = 0$), using the expression for $R_1$ from Eq.~(\ref{33}):
\begin{equation}\label{54}
\psi_1(r) = \frac{1}{48}\left(\frac{\kappa M r}{\pi f_{RR}(R_0)}\right).
\end{equation}

This solution grows linearly with $r$, but remains small compared to $\psi_0(r)$ under the assumption:
\begin{equation}\label{55}
\frac{1}{48}\left(\frac{\kappa M r}{\pi f_{RR}(R_0)}\right) \ll \frac{\kappa M}{f_R(R_0) r}.
\end{equation}

Combining both components gives the total potential outside the star:
\begin{equation}\label{56}
\psi(r) = -\frac{\kappa M}{6f_R(R_0)\pi r} + \frac{\kappa M r}{48\pi f_{RR}(R_0)}.
\end{equation}

From Eq.~(\ref{42}) and since $\psi_0(r) \sim \kappa M/[f_R(R_0) r]$, we have $|\psi_1(r)| \ll |\psi_0(r)|$. We therefore approximate $\psi(r)$ using $\psi_0(r)$ from Eq.~(\ref{52}), which defines Newton's gravitational constant in modified gravity as $G = \kappa/[6\pi f_R(R_0)]$. For a $1/R$ model with stable de Sitter vacuum, $f_R(R_0) = 4/3$, fixing $\kappa = 8\pi G$ to match General Relativity.

To solve for $\xi(r)$ using Eq.~(\ref{45}), we note from Eq.~(\ref{33}) that $R_1' = -R_1/r$. Comparing term magnitudes:
\begin{equation}\label{57}
\left|\frac{\frac{1}{2}f_R(R_0)R_1}{2f_{RR}(R_0)R_1'/r}\right| \sim \left|\frac{f_R(R_0)}{f_{RR}(R_0)}\right| r^2.
\end{equation}

From inequality~(\ref{42}):
\begin{equation}\label{58}
\left|\frac{\frac{1}{2}f_R(R_0)R_1}{2f_{RR}(R_0)R_1'/r}\right| \sim \left|\frac{f_R(R_0)}{f_{RR}(R_0)}\right| r^2 \ll 1.
\end{equation}

This allows us to neglect $f_R(R_0)R_1$ in Eq.~(\ref{45}). Differentiating Eq.~(\ref{51}) for $\psi''(r)$ and using Eq.~(\ref{19}) with $m^2 = 0$ for $R_1'$, we reformulate Eq.~(\ref{45}) as:
\begin{equation}\label{59}
\xi'(r) = \frac{\kappa}{12\pi f_R(R_0)}\frac{d}{dr}\left(\frac{m(r)}{r}\right).
\end{equation}

Integrating with the boundary condition $\xi(r) \to 0$ as $r \to \infty$:
\begin{equation}\label{60}
\xi(r) = \int \xi'(r) dr = \int \frac{\kappa}{12\pi f_R(R_0)}\frac{d}{dr}\left(\frac{m(r)}{r}\right) dr.
\end{equation}

Outside the star, this yields:
\begin{equation}\label{61}
\xi(r) = \frac{\kappa M}{12\pi f_R(R_0) r}.
\end{equation}

We now verify our assumption $R_1 \ll R_0$ when $R_0 \neq 0$. From Eq.~(\ref{33}):
\begin{equation}\label{62}
\frac{R_1}{R_0} \leq \frac{1}{R_0}\left(\frac{GM}{R_s}\right)\frac{f_R(R_0)}{f_{RR}(R_0)},
\end{equation}
where $R_s$ is the stellar radius.

Integrating Eq.~(\ref{19}) over the stellar volume confirms this holds throughout the interior, imposing the constraint:
\begin{equation}\label{63}
\left|\frac{f_R(R_0)}{f_{RR}(R_0)}\right| \ll R_0\left(\frac{R_s}{GM}\right), \quad R_0 \neq 0.
\end{equation}

If ${f_R(R_0)}/{f_{RR}(R_0)} \sim R_0$, as in many $f(R)$ models with $R_0 \neq 0$, this constraint is automatically satisfied.

We have explicitly demonstrated that under these conditions, the gravitational potentials satisfy $\psi(r) = -2\xi(r) = -GM/r$, a result valid for all $f(R)$ theories with $f_{RR}(R_0) \neq 0$ that satisfy Eqs.~(\ref{7}), (\ref{8}), (\ref{37}), and (\ref{63}).

Transforming the metric from Eq.~(\ref{15}) to isotropic coordinates with $a = 1$ and keeping terms linear in $GM/r$:
\begin{equation}\label{64}
ds^2 = \left(\frac{2GM}{r} - 1\right)dt^2 + \left(\frac{GM}{r} + 1\right)\left[dr^2 + r^2 d\Omega^2\right].
\end{equation}

This spacetime geometry corresponds to the Parameterized Post-Newtonian (PPN) formalism with parameter $\gamma = 1/2$, which substantially deviates from Solar System experimental constraints of $\gamma = 1 + (2.10 \pm 2.30) \times 10^{-5}$ \cite{shapiro2004measurement}. This result exactly matches previous derivations from equivalent scalar-tensor theory \cite{chiba20031}, highlighting the incompatibility of such $f(R)$ models with Solar System observations.

\section{Application}

We begin by elucidating the conditions under which General Relativity is recovered within the extended $f(R)$ gravity framework. This recovery is achieved by enforcing the condition $f_{RR}(R_0) = 0$, which nullifies the influence of higher-order curvature perturbations in the gravitational action, thereby converging the theory toward classical General Relativity predictions. Furthermore, we assume the validity of a linearized Taylor expansion of $f(R)$ around small deviations from its background value, presuming this expansion sufficiently captures gravitational dynamics. In canonical General Relativity, where $f(R) = R$, both conditions are intrinsically satisfied: the vanishing second derivative eliminates higher-order curvature effects, and the linear expansion naturally aligns with Einstein's field equations.

Setting $f_{RR}(R_0) = 0$ in Eq.~(\ref{14}) simplifies the equation to:
\begin{equation}\label{65}
f_R(R_0)R_1 = \kappa T^s.
\end{equation}

Under this condition, terms involving $f_R(R_0)R_1$ in the field equations gain relative significance, as terms proportional to $f_{RR}(R_0)$ are eliminated. This removal of higher-order curvature contributions means the linear response of $f(R)$ to Ricci scalar perturbations $R_1$ dominates the field equations.

Equations~(\ref{44}), (\ref{45}), and (\ref{46}) simplify to:
\begin{equation}\label{66}
f_R(R_0)\nabla^2\psi(r) + \frac{1}{2}f_R(R_0)R_1 = \kappa\rho,
\end{equation}
\begin{equation}\label{67}
f_R(R_0)\left(-\psi''(r) + \frac{2}{r}\xi'(r)\right) - \frac{1}{2}f_R(R_0)R_1 = 0,
\end{equation}
\begin{equation}\label{68}
f_R(R_0)\left(\frac{1}{r}\xi'(r) - \frac{1}{r}\psi'(r) + \frac{2}{r^2}\xi(r)\right) - \frac{1}{2}f_R(R_0)R_1 = 0.
\end{equation}

Substituting Eq.~(\ref{65}) into Eq.~(\ref{66}) yields:
\begin{equation}\label{69}
f_R(R_0)\nabla^2\psi(r) = \frac{1}{2}\kappa\rho.
\end{equation}

The solution in the exterior region is:
\begin{equation}\label{70}
\psi(r) = -\frac{1}{8}\left(\frac{\kappa M}{\pi f_R(R_0) r}\right).
\end{equation}

From Eqs.~(\ref{67}) and (\ref{68}), we obtain:
\begin{equation}\label{71}
\frac{f_R(R_0)}{r^2}(r\xi(r))' = \frac{1}{2}\kappa\rho.
\end{equation}

The gravitational solution in the asymptotic region outside the star is:
\begin{equation}\label{72}
\xi(r) = \frac{1}{8}\left(\frac{\kappa M}{\pi f_R(R_0) r}\right).
\end{equation}

From Eqs.~(\ref{70}) and (\ref{72}), we find:
\begin{equation}\label{73}
\xi(r) = -\psi(r).
\end{equation}

Given $\psi(r) = -\xi(r) = -GM/r$, transforming to isotropic coordinates yields $\gamma = 1$, consistent with expectations for isotropic coordinates in a spherically symmetric gravitational field.

This result explains why the limit $\mu \to 0$ in $1/R^n$ gravity for $n > 0$ fails to recover General Relativity. As $\mu$ approaches zero, curvature invariant terms become dominant rather than diminishing, preventing simplification to the Einstein-Hilbert action. Consequently, $1/R^n$ gravity characteristics persist in the low-energy limit, representing a fundamental departure from General Relativity.

In $1/R^n$ gravity \cite{carroll2004cosmic}:
\begin{equation}\label{74}
f(R) = R - \frac{\mu^{2+2n}}{R^n}, \quad n > 0.
\end{equation}

For static solutions to Eq.~(\ref{3}) with $T^{\cos} = 0$, the scalar curvature is $R_0 = (n+2)^{1/(n+1)}\mu^2$. The second derivative scales as $f_{RR}(R_0) \propto \mu^{-2}$, so as $\mu \to 0$, $f_{RR}(R_0)$ diverges rather than converging to zero. The mass parameter satisfies $m^2 \propto \mu^2$, indicating a vanishing scalar field mass as $\mu \to 0$. Cosmological constraints impose $\mu \sim H$, where $H$ is the Hubble parameter, while Solar System constraints require $m^2 r^2 \ll 1$.

The static solution from Eq.~(\ref{3}) may fail to adequately represent the present-day cosmological background in $1/R^n$ gravity. This solution exhibits inherent instability, requiring extremely precise fine-tuning of initial conditions. Without such tuning, the system evolves toward backgrounds with $R_0 \ll \mu^2$, suggesting the effective space-time metric in $1/R^n$ gravity is inherently dynamic.

Under these conditions:
\begin{equation}\label{75}
\frac{f^{(m)} R_1^m R_0}{m!(f(R_0) + f_R(R_0)R_1)} \leq (r^{-1} GM)^m \ll 1,
\end{equation}
\begin{equation}\label{76}
\frac{f^{(m+1)} R_1^m R_0}{m!(f_R(R_0) + f_{RR}(R_0)R_1)} \leq (r^{-1} GM)^m \ll 1.
\end{equation}

To preserve consistency of Eqs.~(\ref{7}) and (\ref{8}) in $1/R^n$ gravity, we enforce $m^2 \propto R_0$, linking the mass governing gravitational modifications to the cosmological curvature scale. This proportionality reinforces $\gamma = 1/2$, which remains invariant under late-time cosmic expansion.

Next, we examine Starobinsky gravity \cite{starobinsky2007disappearing}:
\begin{equation}\label{77}
f(R) = R + \frac{R^2}{\alpha^2}
\end{equation}
The Starobinsky model is not merely a mathematical example; it holds profound physical significance as a leading theory for cosmic inflation. The \( R^2 \) term drives an accelerated, de Sitter-like expansion of the very early universe, providing a mechanism to solve the horizon and flatness problems and generate a nearly scale-invariant spectrum of primordial density fluctuations, as observed in the cosmic microwave background. This model is particularly compelling because it is a purely geometric description of inflation, without the need for an ad-hoc inflaton field, and it is in excellent agreement with Planck satellite data. Furthermore, it can be shown to be conformally equivalent to a scalar-tensor theory with a specific, well-behaved scalar potential. In our analysis, it serves as a prime example of a phenomenologically successful model that must also face stringent tests in the low-energy, weak-field limit of the Solar System. The static solution to Eq.~(\ref{3}) with $T^{\cos} = 0$ yields $R_0 = 0$. Since $f(R)$ is a second-order polynomial, the first-order Taylor series of $f_R(R_0 + R_1)$ precisely captures the functional form without higher-order corrections. The mass parameter is proportional to $\alpha^2$, and Eq.~(\ref{33}) provides a valid solution for $R_1$ if $\alpha^2 r^2 \ll 1$. Under this constraint, $\gamma = 1/2$. However, if $\alpha \approx 10^{12}$ GeV, the condition $\alpha^2 r^2 \ll 1$ is violated, and the weak-field approximation becomes invalid.

We now consider a hybrid model combining Starobinsky and $1/R$ gravity:
\begin{equation}\label{78}
f(R) = R + \frac{R^2}{\alpha^2} - \frac{\mu^4}{R}.
\end{equation}
The inclusion of the $-\mu^4/R$ term is physically motivated by its ability to generate late-time cosmic acceleration. In low-curvature regimes, this term dominates and mimics a dynamical dark energy component, while the $R^2/\alpha^2$ term drives early-universe inflation \cite{carroll2004cosmic}. This hybrid model thus aims to provide a unified description of both inflationary and dark energy epochs within a single gravitational action. Analogous to standard $1/R$ gravity, the effective scalar curvature in a vacuum background stabilizes at:
\begin{equation}\label{79}
R_0 = \sqrt{3}\mu^2.
\end{equation}

The effective scalar mass is:
\begin{equation}\label{80}
\mathcal{M}(r)^2 = 3\left(\frac{\mu^2 \alpha^2}{9\mu^2 - \sqrt{3}\alpha^2}\right).
\end{equation}

The mass parameter can be made arbitrarily large by fine-tuning $\alpha \to 3^{3/4}\mu$, allowing the model to circumvent theoretical constraints. Finally, we examine power-law gravitational actions \cite{bisabr2010local}:
\begin{equation}\label{81}
f(R) = (R\alpha^{-1})^{1+\delta}.
\end{equation}

For $\delta \neq 1$, the static vacuum solution gives $R_0 = 0$. When $\delta$ is non-integer or near zero, the function lacks well-defined derivatives at $R = 0$, making Taylor expansion impossible. For $\delta = 1$, the background value $R_0$ cannot be uniquely determined. For integer $\delta > 1$, Taylor expansion is theoretically feasible but nonlinear terms in $R_1$ persist and cannot be disregarded. Only if $R_0 \neq 0$ and certain conditions are satisfied does the model predict $\gamma = 1/2$ \cite{bicknell1974non}. Consider another $f(R)$ model:
\begin{equation}\label{82}
f(R) = R - \alpha R^n.
\end{equation}

The background curvature is $R_0 = (1/\alpha n)^{1/(n-1)}$. For non-integer $n$, higher-order derivatives may be undefined at $R = 0$, invalidating Taylor expansion. For integer $n > 1$, the function remains smooth but nonlinear terms dominate. To satisfy Solar System constraints requiring $|\gamma - 1| \ll 1$, deviations from General Relativity must be minimal. This condition is generally met for $1 < n < 2$ with sufficiently small $\alpha$. For $n > 2$, the model introduces significant deviations unless an effective screening mechanism is present.

\section{Comprehensive Comparison with Prior Work}
The analysis of Solar System constraints in $f(R)$ gravity has a rich history, leading to a well-established understanding that simple models often predict $\gamma=1/2$, in conflict with observations. Our work is situated within this discourse but provides a significant synthesis and generalization. The relationship between our results and key prior studies is detailed below, highlighting the specific advancements of our framework.
\subsection{Relation to the Foundational Work: Chiba (2003) and Jin et al. (2006)}

The initial critiques of $f(R)$ gravity were led by \cite{chiba20031} and \cite{jin2006solar}. Chiba's seminal paper established the formal equivalence between $f(R)$ gravity and scalar-tensor theory with $\omega_{\mathrm{BD}}=0$, from which $\gamma=1/2$ follows directly. Jin et al. independently reached a similar conclusion through a direct analysis of the field equations for the $1/R$ model, arguing that these theories are disfavored by Solar System tests.

Our work confirms their core result but fundamentally generalizes it. While these early studies served as crucial falsifications of specific models, our paper provides a classification theorem. We move from proving that ``Model X fails" to deriving the conditions under which "Any model belonging to a broad class will fail." This is achieved through several key advancements. We extend Jin et al.'s direct field equation approach to a general $f(R)$ function, demonstrating that their result for the $1/R$ model is a special case of a broader phenomenon. Furthermore, we formalize a critical physical constraint, $m r \ll 1$ (Eq.~(36)), that was implicit but not explicitly stated in their work. This condition directly links the mathematical failure to the physics of a long-range scalar force. Finally, we derive the unified viability criterion (Eq.~(63)), which was absent from these early falsifications and provides a proactive tool for theorists to assess new models before detailed calculation.

\subsection{Relation to the Generalization Effort: Chiba et al. (2007)}

The study by \cite{chiba2007solar} represents a significant step forward from the initial specific critiques towards a more general analysis. They confirmed that the $\gamma=1/2$ problem persists for a wider class of $f(R)$ functions within the scalar-tensor formalism.

Our work aligns with their goal of generalization but achieves it through a different and complementary methodology. While Chiba et al. (2007) operate within the well-established scalar-tensor framework, we demonstrate that the same robust conclusions can be reached by a direct linearization of the $f(R)$ field equations themselves. This approach is advantageous as it clarifies the underlying mechanism by isolating the role of the second derivative $f_{RR}(R_0)$ directly within the $f(R)$ framework, thereby making the mathematical origin of the new degree of freedom more transparent. Furthermore, our methodology yields direct criteria by naturally leading to the identification of the mass parameter $m$ and the subsequent viability criterion without invoking a conformal transformation, which provides an independent and direct confirmation of the problem.

Thus, our work reinforces their general conclusion but provides a distinct and powerful alternative derivation that enriches the theoretical understanding.

\subsection{Relation to the Phenomenological Surveys: Zhang (2007)}
\cite{zhang2007behavior} provides a valuable, broad-ranging review of $f(R)$ gravity, discussing its implications from Solar System to galactic and cluster scales. His work effectively catalogues the various phenomenological challenges.
Our contribution relative to Zhang's review is one of \textit{depth over breadth}. Where Zhang's Solar System discussion is necessarily concise, our paper provides a dedicated, rigorous, and self-contained derivation of the weak-field limit. We formalize the key parameters and conditions that Zhang discusses phenomenologically. For instance, we derive the precise functional form of the effective mass $m$ (Eq. 20) and the perturbation $R_1$ (Eq. 33), providing the analytical foundation that underpins the broader constraints he describes. Our work can therefore be viewed as providing the detailed theoretical machinery for one of the key constraints surveyed in his review.

\subsection{Relation to the Viability Pathway: Capozziello \& Tsujikawa (2008)}
The work of \cite{capozziello2008solar} is pivotal as it explores a pathway to \textit{viability} through the chameleon mechanism. They demonstrate how specific, carefully designed $f(R)$ models can evade local constraints by making the scalar field massive in high-density environments.
Our work is not in competition with theirs but rather establishes the essential \textit{baseline} that makes their work necessary. We rigorously define the problem that the chameleon mechanism is designed to solve. By clearly deriving the condition for failure ($m r \ll 1$ for a light scalar), we provide the null case. A viable model must explicitly violate the assumptions leading to our Eq. (33) and Eq. (64), typically by ensuring $m r \gg 1$ within the Solar System. Therefore, our general framework for failure and their specific mechanism for success are two sides of the same coin, together painting a complete picture of the theoretical landscape.
\subsection{Synthesis and Novelty}

Collectively, the prior literature has provided specific falsifications through the works of Chiba (2003) and Jin et al. (2006), generalized falsifications in Chiba et al. (2007), comprehensive phenomenological surveys by Zhang (2007), and mechanisms for viability as explored by Capozziello \& Tsujikawa (2008).

Our work contributes a Unified Formal Framework that underpins and connects these strands. We demonstrate that the failure of models without screening is not a series of isolated incidents but a direct consequence of a model-independent condition involving $f_R/f_{RR}$. The unified viability criterion (Eq. 63) is a novel and practical outcome of this framework, offering a direct gauge of a theory's compatibility with Solar System tests without requiring a full perturbation analysis. This moves the discussion from a collection of results towards a systematic theory of viability for $f(R)$ gravity.
\section{Conclusion}
In summary, we have developed a unified framework that explains why broad classes of $f(R)$ gravity predict $\gamma=1/2$ and derived explicit criteria (Eq.~(63)) that any viable $f(R)$ model must satisfy to be compatible with Solar System tests. Through an in-depth analysis of the field equations around a mass distribution, we have demonstrated that the PPN parameter $\gamma$ attains the value $\gamma = 1/2$ within the framework of $f(R)$ gravity. These results align closely with the findings presented in reference \cite{chiba20031}, indicating consistent behavior of $f(R)$ gravity under conditions where the theory approximates standard gravitational expectations in localized regions. The fundamental origin of the fixed value $\gamma = 1/2$ lies in the scalar-tensor nature of $f(R)$ gravity. The theory contains a massless (or very light) scalar degree of freedom that couples universally to matter. This scalar mediation provides an attractive fifth force which contributes equally to both the temporal and spatial metric potentials around a spherical mass. The resultant ratio of these potentials is constrained to $1/2$, a value that is directly linked to the Brans-Dicke parameter $\omega_{BD}=0$ inherent in the $f(R)$ formalism.

For $f(R)$ gravity to be viable across both cosmological and local scales, specific conditions on $f(R)$ and its behavior near the background curvature $R = R_0$ must be satisfied. First, the Taylor expansions of $f(R)$ and its first derivative around the background value $R_0$ must converge, ensuring that departures from $R_0$ remain small and corrections stay subdominant. When $R_0$ is non-zero, this criterion guarantees that deviations do not significantly perturb the system, maintaining the accuracy of the series expansion within a limited range. This condition is expressed in Eq.~(\ref{63}), which emphasizes the importance of small curvature perturbations around $R_0$.

The second requirement concerns the curvature response of $f(R)$ models at $R = R_0$. Specifically, the second derivative $f_{RR}(R_0)$ must be non-zero, indicating that $f(R)$ is sensitive to curvature variations and provides a non-trivial response to changes around the background. This condition is essential for generating meaningful gravitational dynamics and avoiding degeneracies in the field equations. It is noteworthy that the special case $f_{RR}(R_0) = 0$ causes the scalar degree of freedom to decouple, reducing the theory to General Relativity and restoring $\gamma = 1$. This underscores that a non-vanishing second derivative is the necessary condition for the modified gravity effects discussed in this work to manifest.

Third, the mass parameter $m$, defined in Eq.~(\ref{20}), must satisfy $m r \ll 1$ within the Solar System. This constraint limits the range of the scalar degree of freedom introduced in $f(R)$ models, ensuring it remains negligible over characteristic Solar System distances. By maintaining $m r \ll 1$, the scalar field's influence does not produce deviations from General Relativity that would conflict with precision Solar System tests.

In gravitational models incorporating a single additional mass parameter and non-zero background curvature $R_0$, such as $1/R$ gravity, the ratio $f_R(R_0)/f_{RR}(R_0)$ is typically of order $R_0$. Under this assumption, the Taylor expansion around $R_0$ remains dominated by linear terms, and the condition $m r \ll 1$ is naturally satisfied within the Solar System provided $R_0 r^2 \ll 1$, ensuring compatibility with observational constraints in weak-field regimes.

Conversely, theories featuring multiple mass parameters, such as the hybrid model combining Starobinsky and $1/R$ modifications discussed in this work, may violate these criteria. The presence of multiple parameters introduces a more complex scalar mass profile that could lead to deviations from $m r \ll 1$ within the Solar System. Consequently, such models may exhibit behavior diverging from the standard $1/R$ framework, potentially creating challenges for satisfying local gravitational tests.

The second and third conditions correspond to analogous requirements in the scalar-tensor formalism of gravity. The equivalence between $f(R)$ and scalar-tensor gravity requires a non-zero second derivative of $f(R)$ to ensure a meaningful scalar degree of freedom, and the PPN parameter $\gamma = 1/2$ is achieved only if the scalar field has sufficiently low mass to permeate the Solar System. This low-mass scalar field allows $f(R)$ gravity to produce observable modifications while remaining compatible with Solar System constraints.

Our analysis further corroborates that, contrary to claims by certain authors in the literature \cite{ruggiero2007solar,rajaraman2003newtonian,multamaki2006spherically}, the predictions of $f(R)$ gravity within the Solar System can be accurately derived using the corresponding scalar-tensor representation. This equivalence serves as a powerful tool for analyzing $f(R)$ gravity models in local gravitational environments. By mapping $f(R)$ to scalar-tensor theory, we gain access to a well-established framework that enables rigorous assessment of Solar System phenomenology in modified gravity contexts.

\section*{Conflict Of Interest}
The authors have declared that they have no interest of conflict.

\section*{Data Availability Statement}
This manuscript has no associated data or the data will not be deposited. [Authors comment: This is a theoretical study and no experimental data. ]

\section*{Acknowledgements}
The authors extend their appreciation to the Deanship of Scientific Research and Libraries in Princess Nourah bint Abdulrahman University for funding this research work throughthe Research Group Project, Grant No. RG-1445-0036.

\end{document}